\documentclass[aip,twocolumn,amsmath,amssymb,reprint]{revtex4-1} \synctex=1 
\usepackage{graphicx}% Include figure files
\usepackage{dcolumn,array}% Align table
\usepackage[colorlinks=true,linkcolor=blue]{hyperref}
\usepackage{hyperref}
\usepackage{lineno}
\usepackage{bm}% bold math
\begin{document}

%\preprint{APS/123-QED}

\title{Surface coating induced lubrication in flowing granular
  materials}

\author{Sayali V. Chaudhary} \author{Ashish V. Orpe}
\email{av.orpe.ncl@csir.res.in}
\affiliation{CSIR-National Chemical Laboratory, Pune 411008 India}
\affiliation{Academy of Scientific and Innovative Research (AcSIR),
  Ghaziabad 201002 India}

\date{\today}% It is always \today, today,
% but any date may be explicitly specified

\begin{abstract}
  We investigate the flow of spherical, bulk granular particles down
  an inclined plane mixed with small-sized spherical lubricant
  particles using discrete element method simulations. Predefined cohesive interaction is implemented between
  lubricant and bulk particles, enabling the coating of the former over the
  latter. The overall flow rate exhibits non-monotonic dependence on
  lubricant content. Initially, it increases with lubricant addition,
  reaches a maximum at an intermediate lubricant content, and
  decreases for even higher lubricant content.  The increase in the
  flow rate is attributed to a lower inter-particle friction coefficient
  between lubricant-coated bulk particles. The decrease in the flow
  rate at higher lubricant content, on the other hand, is attributed
  to enhanced densification and increased damping between crowded
  particles. Both these occurrences are examined using various flow
  level characteristics.  The simulation results are found to be in
  qualitative agreement with previous experimental results. Overall,
  the outcome integrates novel computational insights and prior
  experimental results to enhance the understanding of the powder
  lubrication phenomena.
\end{abstract}

%\pacs{45.70.Mg,47.57.Gc}% PACS, the Physics and Astronomy
% Classification Scheme.
% \keywords{Suggested keywords}%Use showkeys class option if keyword
% display desired
\maketitle

\section{\label{intro}INTRODUCTION}

One of the most important characteristics governing the flow of
granular material is the friction between particles and between
particles and boundaries. This frictional resistance arises from the
sliding or rolling of particles against other surfaces and is an
inherent feature of a granular particle. Investigation of the effect
of friction on granular flow is possible by using particles of
different types of materials, each exhibiting distinct frictional
characteristics. \cite{CarrigyAOR, 2DSilo,
  NiuRingShear} Alternatively, the effect of friction can be
investigated through discrete element method (DEM) simulations
\cite{silbertgranular01, rotatingDEM, siloDEM} to model desired
frictional behavior, but that complicates suitable experimental
validation. A simpler approach to understand friction effects is to
coat the surface of granular particles with other tiny, frictionless
(lubricant) particles.  Indeed, such modification of particle surface
properties is well practiced in pharmaceutical industries for various
applications. \cite{PharmaCoating, Dave_improvedFlow_coating,
  MechDryCoating}

The individual lubricant particles, with fractional size of the bulk
particles, when added in tiny quantities, adhere to the surface of the
latter. Due to their own frictionless characteristics, they are
expected to reduce the friction between bulk particles, thereby
enhancing the overall flowability. Many of the previous scientific
studies~\cite{llusa05,navaneethanapplication,wangcontrolled,ketterhagen18,danisheffect}
on this lubrication phenomenon have mostly been directed toward
understanding the optimality of lubricant mixing, particle size ratios,
and lubricant concentration with respect to overall flow
characteristics. A few studies, however, have attempted to understand
the mechanism and fundamentals of the lubrication process. This
includes rheometry to investigate the flow behavior,~\cite{zhoueffect}
scanning electron microscopy (SEM) to observe surface morphology of
large particles on lubricant coating and its effect on angle of
repose.~\cite{morineffect} A detailed analysis of the flow of a mixture
of lubricant and large particles down an inclined plane has also been
carried out recently.~\cite{ghodakeflow,RavindraparticleEffect} The
above experimental studies mostly reveal the effect of lubricant
content on altered flow properties.~\cite{ghodakeflow,morineffect}  In
a recent study, it was shown that the lubrication effect is found to
be dependent on the flow type, flow duration and system
confinement.~\cite{RavindraparticleEffect}

All the previous studies provide macroscopic information about the
lubricant-induced flow with minimal information about the dynamics
occurring at the particle level in terms of contacts, packing fractions, and surface coverage. Certain information about flow in terms of velocity
profiles has been possible, though only near the
wall.~\cite{ghodakeflow} The understanding of the lubrication process
will, however, require details about the spreading of lubricant over bulk
particles, distributions of spreading, particle contact details, volume
fraction and velocity measurements. These particle-level studies can
be implemented using discrete element method (DEM) simulations. The
main advantage of this method is the ability to study time-dependent
behavior in three dimensions while acquiring all possible details about
lubrication mechanisms, their spreading and coating ability, and the
influence on overall macroscopic flow behavior. Moreover, this method has been used substantially for studying granular systems~\cite{silbertgranular01,Cleary_DEM} and has also been validated
extensively through experiments.~\cite{gdrmididense,RapidGranFlow}
We note one attempt at understanding the lubrication process using this
method.~\cite{yoshidadem} Given the initial attempt at modeling a
complicated particle system comprising mixture of two different sizes
and cohesive interactions, certain limitations exist in terms of
replicating actual experiments. At the same time the study allows for
a detailed understanding of the mechanics of lubrication and provides
a suitable framework for future studies.

\section{\label{method}METHODOLOGY}

\subsection{\label{system-details}Simulation system}

\begin{figure}
  \includegraphics[scale=0.62]{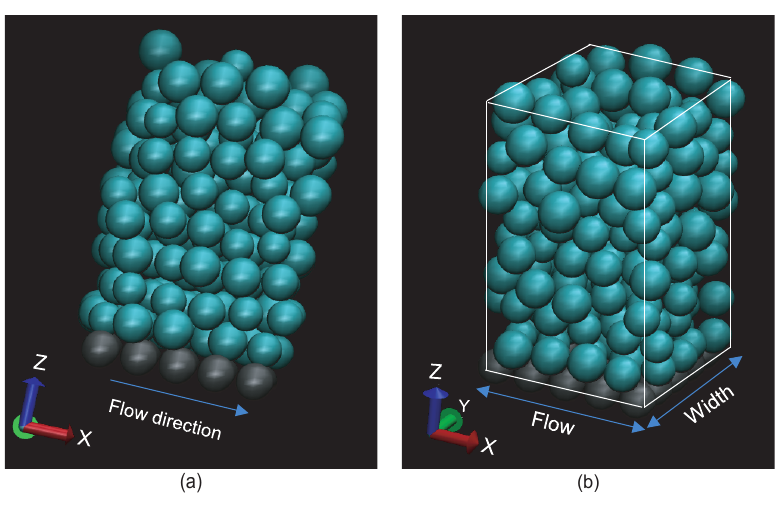}
  \caption{Schematic showing the inclined chute configuration and the
    bulk particles. (a) View from the side. (b) Perspective view
    exhibiting all three directions. The blue spheres represent bulk
    particles, and the gray spheres denote the rough bottom surface. See text for more details.}
  \label{fig1}
\end{figure}

The discrete element method (DEM) simulations were carried out using the
open-source program LIGGGHTS (Lammps Improved for General, Granular
and Granular Heat Transfer Simulations). The simulations executed the
flow of a mixture of two types of granular particles down an inclined
plane. One of them is a spherical (bulk) particle of diameter $d_{p}$
with $15$\% polydispersity in size, and the other is a spherical
(lubricant) particle of diameter $d_{l}$. The simulation geometry, as
shown in Fig.~\ref{fig1}, consists of a rectangular box of length
$5 d_{p}$, width $5 d_{p}$ and height $7 d_{p}$. The flow occurs in
$x-$direction, the transverse to the flow is in $y-$direction, and height
is in $z-$direction. The chute base is made rough by fixing spheres of
the same size and type as the bulk spheres, arranged in a simple cubic
lattice. The top is kept open, allowing for flow expansion/contraction,
typical of a free surface flow. The chute is maintained inclined at
$24^o$ to the horizontal in all the cases studied.

Periodic boundary conditions are applied in the $x-$ and
$y-$directions. Thus, particles exiting from one end of the box in the
$x-$direction are reinserted at same $y$ and $z$ positions at the other
end of the box in $x-$direction. Similar reinsertion also is carried
out in the $y-$direction. This arrangement ensures a constant number of
flowing particles in the box at all times and the occurrence of flow in the central region of the chute away from the side walls and chute
entry-exit. Furthermore, this arrangement allows study of continuous flow
over a very long duration in a smaller system using a finite number of
particles, needed to obtain better statistics. However, these
advantages are accompanied by certain caveats. For instance, an
experimental system can induce particle segregation due to the presence of
walls which does not get reflected in simulations with periodic
boundaries. Thus, the flow in simulations appears to be highly
homogeneous, which may not occur in a finite size experimental
system. Furthermore, the lubricant particles can escape from lower
boundaries and are artificially recirculated in the system (see figure
for flow configurations) to preserve the number of particles. Such an
effect is not observed in experiments. The simulations results, thus,
need to be understood in view of these effects introduced by the use of
periodic boundaries.

As discussed later, the simulation results are compared with previous
experimental data.~\cite{ghodakeflow,RavindraparticleEffect} The
experimental system used a mixture of spherical glass beads
($d_{p} = 2$ mm) as bulk particles and magnesium stearate (MgSt) as
lubricant particles. These lubricant particles are cohesive,
plate-like particles with an effective diameter ($d_{l}$) of $10$ $\mu\text{m}$. These plate-like particles, after mixing with bulk particles,
tend to spread over the latter's surface in the form of a film. The
friction between bulk particles is reduced due to contact between
these lubricant films, which are nearly frictionless with respect to
each other.  However, this reduced frictional property of lubricant
particles is manifested only on spreading and stretching of plate-like
particle over another surface, but not for a separate, standalone
assembly of unstretched lubricant particles. This behavior is quite
well known in pharmaceutical powder mixing,~\cite{morineffect,Li_Lubricants,MixerType_coating} wherein the
lubrication function of MgSt powder is associated with the ability of its particles to coat host particles.

The use of spherical lubricant particles in simulations is a very
simplified approximation of the plate-like lubricant particles used in
experiments. The spherical approximation is implemented purely from
the perspective of reduced computational effort. The objective here is
to ensure a coating of spherical lubricant particles over the surface
of bulk particles, mimicking a film of lubricant as in
experiments. This requires the spherical lubricant particle to be
quite small compared to bulk particles.  As an approximation, the
simulations incorporate a size ratio ($S = d_{p}/d_{l}$) of $10$
between two particles.  While this is small compared to $200$ used in
experiments, it is still large enough so that the surface of bulk
particles can be coated adequately.  It was found that any further
increase in the size ratio led to an increased number of lubricant, hence a total number of particles, thereby increasing the computational cost
substantially. On the other hand, reducing the size ratio made the
lubricant particle size approach the bulk particle size thereby
rendering the lubricant coating ineffective. Furthermore, the lubricant (MgSt)
particle is quite soft and flaky in nature, and its stiffness/modulus is
not reported in the literature. The spherical lubricant in simulations
is, then, approximated to be of the same stiffness as the bulk particle,
typically made from glass. Table~\ref{tab} provides a comparison of
different system and particle-level parameters used in experiments and
simulations.

\begin{table*}
  \caption{\label{tab} Comparison of key setup parameters between simulations and reference experiment (in-house system and materials)}
  \begin{ruledtabular}
    \begin{tabular}{lll}
    \textbf{Parameter} & \textbf{Simulation} & \textbf{Experiment} \\
    \hline
      Chute inclination to horizontal & $24^o$ & $24^o$ \\
      Flow region for analysis & $5d_p \times 5d_p \times$ height ($z$) & $300d_p \times
                                                     100d_p$ (near
                                                                          sidewalls) \\
      Chute width & $5d_p$ (Periodic) & $25d_p$ (with sidewalls) \\
      Bulk particles ($d_{p}$) & Spherical & Spherical (Glass) \\
      Lubricant particles ($d_{l}$)  & Spherical & Plate-like (MgSt) \\
      Particle size ratio ($d_p/d_l$) & 10 & 200 \\
      Particle density ratio ($\rho_p/\rho_l$) & 2.4 & 2.4 \\
      Lubricant specific surface area & $0.58$ $m^{2}/g$ &
                                                                 $1.3 - 10.5$
                                                                 $m^{2}/g$
    \end{tabular}
  \end{ruledtabular}
\end{table*}

\subsection{\label{system-details}Simulation model}

The contact force between any two particles in the system is
calculated using the Hertzian Contact model. The interaction force in this
model is dependent on the overlapping area of interacting particles
and captures the realistic behavior of granular materials quite
well.~\cite{thorntonfrictional,silbertgranular01,silbertgranular03}
The contact force comprises normal ($\bm{F_{n}}$) and tangential
($\bm{F_{t}}$) components, given as,
\begin{equation}
  \bm{F_{n}} = k_{n} \delta \bm{n} - {\gamma_{n}
    \bm{v}_{n}}
\end{equation}
\begin{equation}
  \bm{F_{t}} = k_{t} \Delta s_t - {\gamma_{t}
    \bm{v}_{t}}
  \label{contactforce}
\end{equation}
where $\bm{n}$ is the unit vector normal along the line connecting
the centers of those two particles, $\delta$ is the overlapping distance
between them, $\bm{v}_{n}$ and $\bm{v}_{t}$ are the normal and
tangential components of particle velocities, and $\Delta s_t$ is the
tangential displacement vector between two particles. Static friction
is applied via the Coulomb yield criterion ($\bm{F_{t}}$ $\le$
$\mu_s\bm{F_{n}}$). All parameters are non-dimensionalized such that
the bulk particle has diameter $d_{p}$ = 1 and mass $m_{p}$ = 1 with
gravity (of magnitude $g = 1$) acting downward in the $z $-direction.
The simulations use a natural time unit as $\tau = \sqrt{d_p/g}$. 

The lubricant particles adhere to the surface of bulk particles
through cohesive force, modeled using the simplified
Johnson-Kendall-Roberts (SJKR) framework. This approach introduces a
normal, cohesive force between two lubricants and bulk particles in
contact, in addition to the existing Hertzian contact
force~\cite{roesslerdem,ramirezaragoncomparison} and is given as
\begin{equation}
  \bm{F_{cohesion}} = C A
  \label{cohesion}
\end{equation}
where $A$ represents the contact area between interacting particles
and $C$ is the cohesion energy density representing the strength of
attraction between lubricant and bulk particles. The value of C is set
to be $8 \times 10^{4} m_pg/d_p^2$, which is large enough
($> 4 \times 10^{4}$) to enable sticking of lubricant particles to
bulk spheres but also small enough ($< 12 \times 10^{4}$) to prevent
excessive agglomeration between particles causing plug flow in the
system, never observed in experiments.~\cite{ghodakeflow}  The exact
dependence of flow and lubrication on cohesion energy density,
however, was not investigated, as that was out of the purview of this
work. At the same time, it is expected that the overall flow behavior
would remain qualitatively the same for all values of cohesion strength
within the range specified above. The value of cohesion energy density
considered here and the ensuing normal force results in bulk-lubricant
particle overlap of $0.5$\% of the effective particle radius, which is
within the advisable range to prevent unrealistic flows,~\cite{Cleary_overlap, Stiffness-overlap_review} It is to be noted
that no cohesive force was considered between two lubricant particles,
as that led to the formation of clustering not observed in
experiments.~\cite{ghodakeflow,RavindraparticleEffect} For
simulations of only bulk particles (no lubricant and cohesive force),
a time step of $6 \times 10^{-5}$ is employed, whereas for simulations
of lubricant added system (cohesive forces), a smaller time step of
$1 \times 10^{-5}$ is used. This corresponds to $\Delta t = 0.065t_H$,
where $t_H$ is the Hertzian contact time, which lies well below the
commonly recommended numerical stability limit of $0.1-0.2$ $t_H$
used in DEM simulations.

The initial state of particles is achieved by pouring a mixture of a
predefined number of bulk and lubricant spheres in the simulation box,
which is bounded by vertical walls in $x-$plane. The box is
maintained inclined at $24^o$ to the horizontal. During pouring,
the lubricant particles adhere to the bulk spheres due to cohesive
interactions. The filling is carried out up to a height of $7 d_{p}$
and particles are allowed to settle properly in the box following
which the vertical walls are removed to initiate the flow.  The
simulation progresses in time leading to flow evolution and
redistribution of lubricant particles over bulk particles
surface. Simulations are carried out till steady state is reached,
marked by constant flowing layer thickness and the total kinetic
energy remaining constant to within $1$\% of the mean value. All
subsequent analysis is performed in the steady-state flow regime
spanning over a time duration corresponding to a mean flow travel
distance of at least $2000 d_{p}$.

The primary focus of this work is to investigate the effect of powder
lubricant on the flow behavior of bulk particles and the underlying
lubrication mechanism.  The concentration of lubricant in the granular
mixture is quantified as number ratio ($N_r$), defined as the ratio of
the number of lubricant particles to that of bulk
particles. Simulations are conducted for a wide range of number ratios
($N_{r}$).  For each value of $N_r$, the instantaneous velocity of the
particles is recorded at regular time intervals throughout the
simulation. The flowing layer is divided into bins parallel to the
flow direction, each with length $5 d_{p}$ and thickness $1 d_p$. The
flow direction average velocity in each bin ($z-$ coordinate) is calculated as
$v_{x} = \langle c_{x} \rangle$, where $c_{x}$ is the
instantaneous particle velocity in flow ($x-$) direction. The number
averaging ($\langle . \rangle$) is done over all particles located
within the bin at any given instant, followed by averaging these
values over $1000$ time intervals at steady state. The value of
$v_{x}$ is determined separately for bulk, lubricant and particle
mixture. The variation of $v_{x}$ with height ($z-$ coordinate)
provides the flow direction average velocity profile.  The mean flow rate
$\langle Q \rangle$ in the flowing layer is calculated as the product
of mean volume fraction, flow width ($5 d_{p}$), flowing layer
thickness ($z$) and flow depth average velocity
$\langle v_{x} \rangle$ after subtracting bed slip velocity ($v_{s}$).

\subsection{\label{parameter}Simulation parameters}

\begin{figure}
  \includegraphics[scale=0.45]{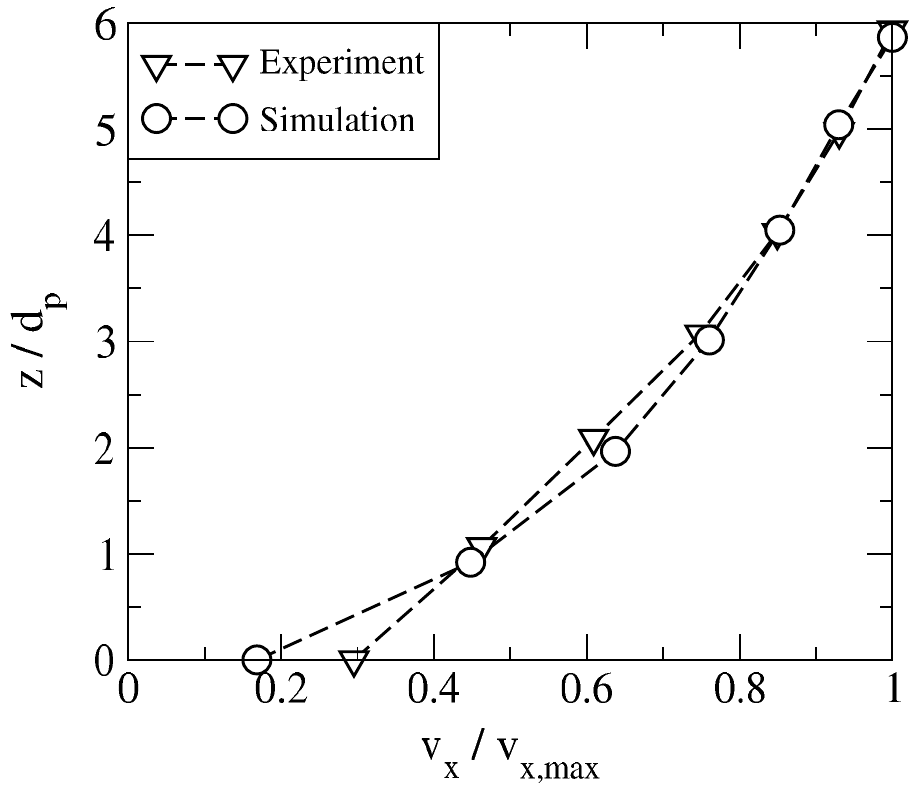}
  \caption{Variation of steady-state, normalized, flow direction average
    velocity ($v_{x} /v_{x, max}$) profile with normalized distance from the chute bed ($z/d_{p}$) in simulations and experiments (unpublished).
    $v_{x.max}$ represents the maximum velocity in the flowing layer
    at the free surface and $d_{p}$ denotes the bulk particle
    diameter. See text for more details.}
  \label{fig2}
\end{figure}

\begin{table}
  \caption{\label{tab-1} Values of coefficient of friction ($\mu$) and
    restitution ($e$) for different pairs of particles used in simulations}
  \begin{ruledtabular}
    \begin{tabular}{lccc}
      Bulk inter-particle friction coefficient & $\mu_{pp}$ & 0.30 \\
      Bulk-bed friction coefficient & $\mu_{pb}$ & 0.20 \\
      Bulk-lubricant friction coefficient & $\mu_{lp}$ & 0.10 \\
      Lubricant inter-particle friction coefficient & $\mu_{ll}$ &
                                                                   0.01 \\
      Lubricant-bed friction coefficient & $\mu_{lb}$ & 0.10 \\ 
      Bulk inter-particle restitution coefficient & $e_{pp}$ & 0.70 \\
      Bulk-bed restitution coefficient & $e_{pb}$ & 0.70 \\
      Bulk-lubricant restitution coefficient & $e_{lp}$ & 0.30 \\
      Lubricant-bed restitution coefficient & $e_{lb}$ & 0.30 \\
      Lubricant inter-particle restitution coefficient & $e_{ll}$ &
                                                                    0.05 \\
    \end{tabular}
  \end{ruledtabular}
\end{table}

The simulation model comprises several parameters, which need to be
calibrated properly, given two different types of materials, to
generate realistic flow. The foremost among all the parameters is the value of the friction coefficient ($\mu$), which is expected to be the
primary driver in altering the flow behavior and has been calibrated
to the best possible extent. The tangential ($k_{t}$) and normal ($k_{n}$)
elastic constants for bulk particles are set to $O(10^6 m_pg/d_p)$, as
used previously,~\cite{silbertgranular01,zhanginclined,gdrmididense}
which are appropriate for glass beads.  The same values were also
applied for lubricant particles, for which these parameters are not
readily available.  The values of tangential ($\gamma_{t}$) and normal
($\gamma_{n}$) damping constants were selected to obtain the
coefficient of restitution ($e$) shown in Table~\ref{tab-1} for
different interacting particle pairs. To calibrate friction coefficients,
we simulated the flow of bulk particles in a chute setup, inclined at
an angle of $24^o$ and comprising a box with periodic boundaries
in the $x$-direction along with flat side walls and a rough bottom surface.
The length of the periodic box was $10 d_{p}$, while the width was
$25 d_{p}$ bounded with flat walls. The bottom was made rough using
bulk particles arranged in a cubic lattice with an open top surface.

Simulations in this inclined chute system were initially executed for
the flow of bulk spheres in the absence of lubricant particles. The
flow layer velocity profile was measured near one of the sidewalls as
was also done in experiments with flowing glass beads.~\cite{ghodakeflow}  The values of friction coefficients for
inter-particle contacts ($\mu_{pp}$), particle-sidewall contact
($\mu_{pw}$) and particle-rough bottom contacts ($\mu_{pb}$) were
suitably adjusted to match the experimentally obtained profile. The
value of $\mu_{pb}$ is set slightly lower than $\mu_{pp}$ to account
for the slip at the bottom surface observed in experiments. The
comparison between normalized velocity profiles from simulations and
experiments is shown in Fig.~\ref{fig2} for the optimized values of
friction coefficients. The matching looks reasonably close, except for the
slip velocity, suggesting that the chosen values of friction
coefficients may provide realistic flow behavior.

\begin{figure*} 
\includegraphics[scale=0.55]{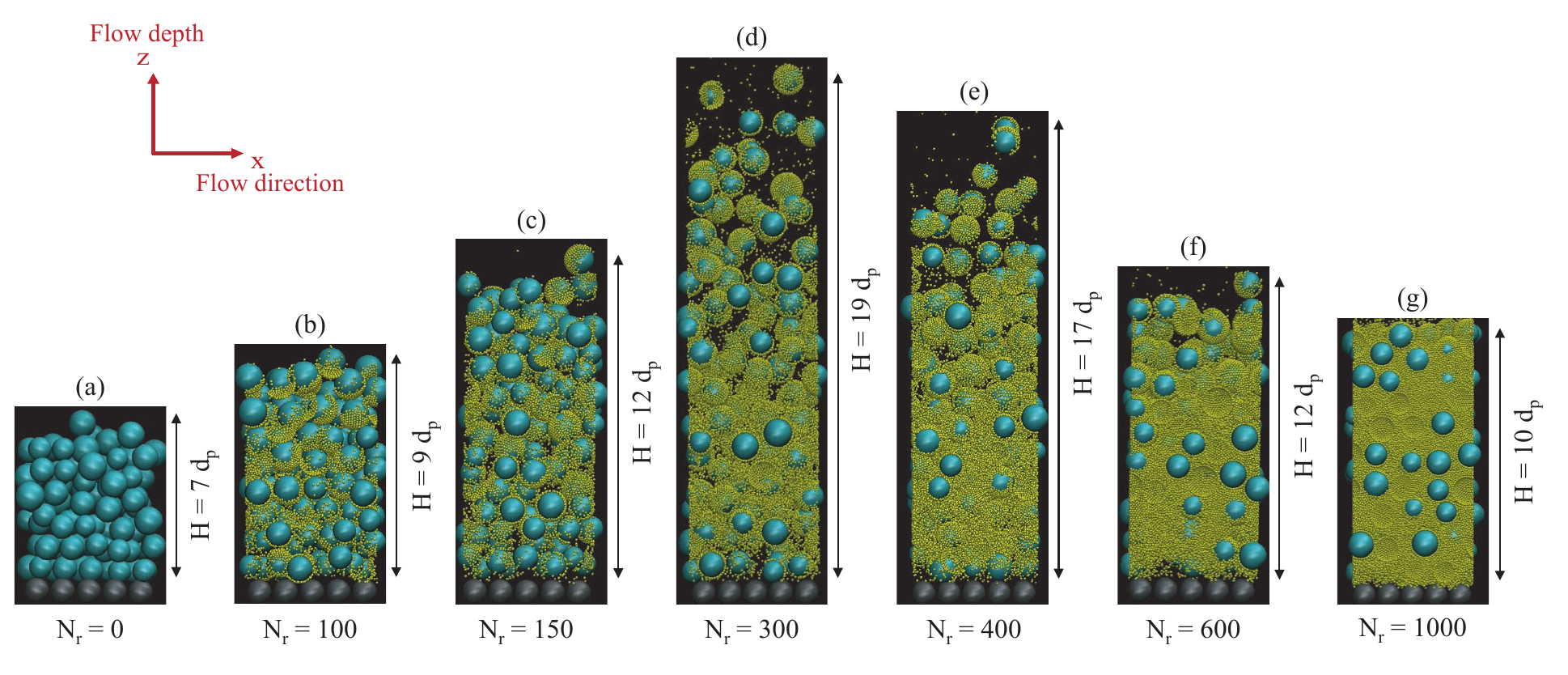}
  \caption{Instantaneous configurations (images) from simulations as viewed from the side at steady-state for different values of ($N_{r}$). Bulk particles are represented as large blue spheres and lubricant particles as small yellow spheres. Cohesive interaction between bulk and lubricant particles is as per eq.~(\ref{cohesion}). Multimedia available online}
  \label{fig3}
\end{figure*}

These calibrated parameters for bulk particles were then used to
explore the lubricant-induced flow of bulk particles. For lubricant
particles, the inter-particle friction coefficient ($\mu_{ll}$) is
assumed to be 0.01 to qualitatively capture lubrication effects. It is
not intended to represent an exact value of material property, but
rather to represent nearly frictionless interfaces of lubricant
particles occurring after coating.~\cite{millerfrictional1985} The
restitution coefficient between two lubricant particles
($e_{ll} = 0.05$) is based on exact value from a previous
study~\cite{serrisdry2013} reflecting dissipative nature of
lubricant. The values of the coefficient of friction between lubricant and
bulk particles ($\mu_{lp}$) and between lubricant and bed particles
($\mu_{lb}$) were assumed to lie between $\mu_{pp}$ and
$\mu_{ll}$. Exact values were indeterminable since there was no
experimental system to compare and calibrate the coefficients. The
value of lubricant-sidewall friction coefficient ($\mu_{lw}$) was not
required to be determined since the simulations were carried out under
periodic boundary conditions. The finalized values of coefficients of
friction and restitution are listed in Table~\ref{tab-1}.

\section{\label{results}RESULTS AND DISCUSSION}

We first discuss the qualitative changes in the flow morphology and
layer expansion with varying lubricant content. These changes are then
quantified in terms of varying packing fraction, velocity profiles, and
flow rate. The details of lubricant coating are described
thereafter. Finally, we discuss the role of cohesion between lubricant
and bulk particles influencing the overall lubrication process.

\subsection{\label{parameter}Flow morphology and packing fraction}

A set of instantaneous images of the steady-state flowing layer, as
seen from the side, is shown in Fig.~\ref{fig3} (Multimedia view) for varying lubricant content ($N_{r}$). The accompanying
supplementary material provides relevant details about all the videos.
A non-monotony is clearly evident from these images. The particles are
densely packed in the flowing layer in the absence of lubricant particles
($N_{r} = 0$).  This flowing layer, then, progressively expands with
increasing lubricant content till maximum layer thickness is observed
at $N_{r} = 300$ and thereafter progressively contracts for further
increase in $N_{r}$. It is to be noted that the extent of expansion is
considerably more than the actual increase in total particle volume
from added lubricant spheres. For instance, the layer thickness for
$N_{r} = 300$ is more than twice that for $N_{r} = 0$ even though the
actual increase in total particle volume is only $30$\%. The
exaggerated expansion can be thought to arise due to increased
collisions from faster flow. At the same time, this expansion results
in dilation thereby assisting improved flowability.  However, this
effect seems to reverse for values of $N_{r}$ beyond $300$, wherein
the flow contracts leading to densification. This may be related to
lower flow velocity and reduced collisions or increased damping. A
qualitatively similar non-monotonic variation of flowing layer
thickness with respect to lubricant concentration has been observed
previously in experiments.~\cite{ghodakeflow,RavindraparticleEffect}

\begin{figure}
  \includegraphics[scale=0.45]{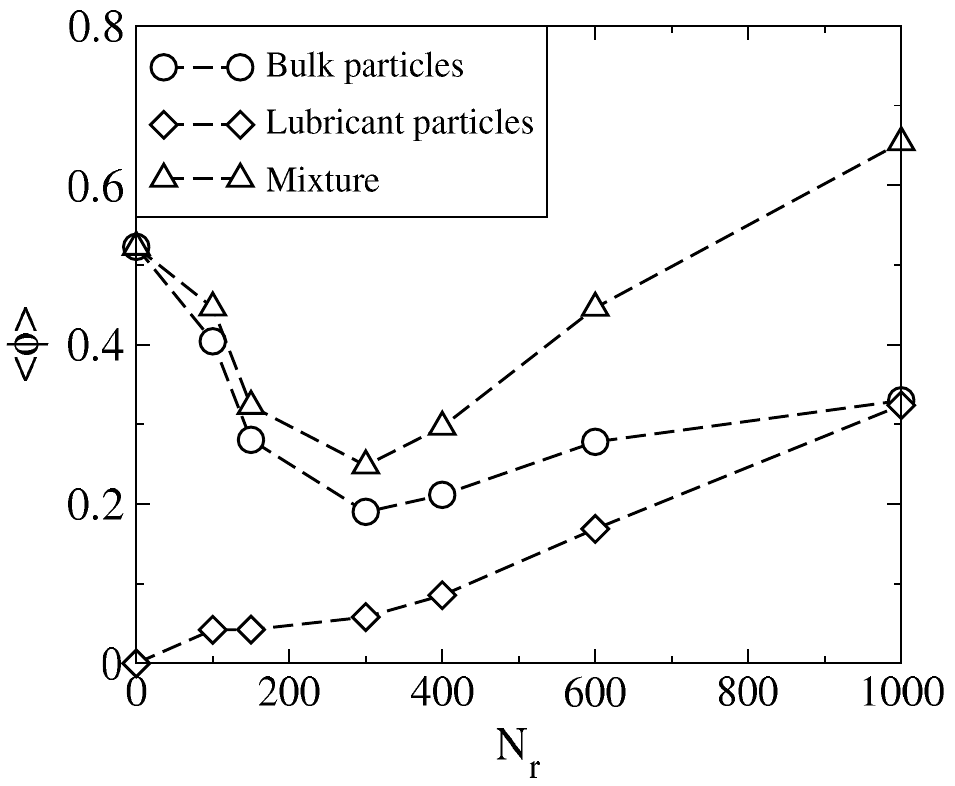}
  \caption{Variation of steady-state mean volume fraction ($\langle \phi \rangle$) of particles (bulk, lubricant, and mixture) in the flowing layer with the number ratio ($N_{r}$). Error bars are smaller than the symbol size and hence are
    not shown to ensure clarity in the figure. See text for more details.}
  \label{fig4}
\end{figure}

The flow expansion and compression observed in Fig.~\ref{fig3} is
quantified in terms of particle volume fraction ($\phi$) variation with $N_{r}$.  Detailed profiles of $\phi$ in the flowing layer are shown in supplementary material Fig.S1 for different values of $N_{r}$. Over here, the mean values ($\langle\phi\rangle$) are shown in Fig.~\ref{fig4} for bulk particles, lubricant
particles, as well as their mixture. The volume fraction for bulk particles as
well as the mixture exhibits a non-monotonic dependence on $N_{r}$,
though with certain differences. The decrease in bulk particle volume
fraction to a minimum is attributed to the expansion phase of the
flowing layer [see Figs.~\ref{fig3}(a)-\ref{fig3}(d)]. Correspondingly, the volume
fraction of mixture also decreases in this phase. However, the values
are consistently higher than those for the bulk phase due to the volume of
lubricant particles increasing monotonically (diamonds in
Fig.~\ref{fig4}), which occupy the void
spaces between bulk particles. It may be reasoned that the presence of
lubricant particles disrupts the tight packing of bulk particles by
reducing inter-particle friction and enhancing mobility, leading to a
more loosely packed and expanded flowing layer.

During the compression phase of the flowing layer [Figs.~\ref{fig3}(d)-\ref{fig3}(g)]
and data for $N_{r} > 300$ in Fig.~\ref{fig4}), the bulk particles
volume fraction does not increase appreciably. This is due to the
continuously increasing number of lubricant particles with $N_{r}$,
which occupy more space and prevent bulk particles from coming closer
to each other and almost forever keeping them in the dilute state. On
the other hand, the mixture volume fraction increases much faster. In
fact, at $N_{r} = 1000$, the volume fraction of the mixture is larger
than the volume fraction for bulk particles only ($N_{r} = 0$)
indicating improved packing efficiency. This is attributed to the
continuous addition of lubricant particles, which increased their volume
fraction while occupying the space between bulk particles. Similar,
qualitative non-monotonic variation of packing fraction with lubricant
concentration has been observed previously in
experiments,~\cite{ghodakeflow,RavindraparticleEffect} but with a key
difference. The packing fraction of particles in the flowing layer in
experiments is primarily due to the packing of bulk particles since a
very tiny, non-measurable volume is occupied by the small-sized
lubricant particles. This is unlike the substantial volume occupied by
lubricant particles as observed in simulations, especially at higher
values of $N_{r}$, which seems to dominate overall packing
fraction. This gets manifested more in the data for flow rates and
velocities as described next.

\subsection{\label{parameter}Flow rate and velocity profiles}

\begin{figure}
  \includegraphics[scale=0.45]{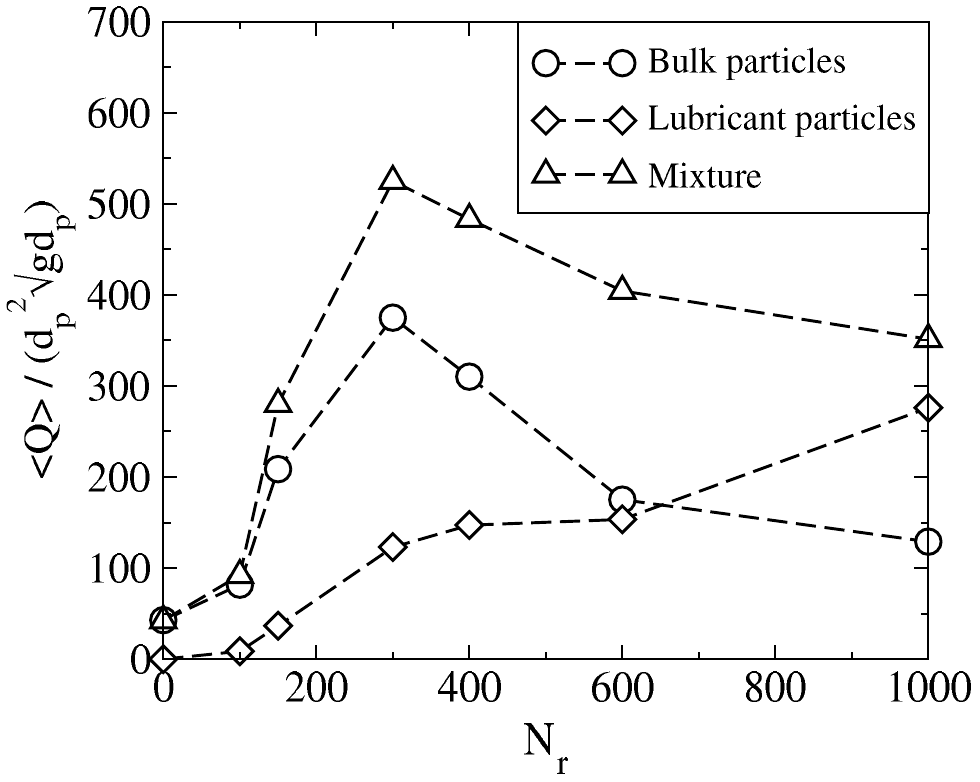}
  \caption{Variation of steady-state, normalized, mean flow rate
    ($\langle Q \rangle/d_{p}^{2}\sqrt{g
  d_{p}}$) of particles (bulk, lubricant, and mixture) in the flowing layer
  with the number ratio ($N_r$). Error bars are smaller than the symbol size and hence are not shown to ensure clarity in the figure. See text for more details.} 
  \label{fig5}
\end{figure}

The variation of mean volumetric flow rate ($\langle Q \rangle$) with
$N_{r}$ is shown in Fig.~\ref{fig5} for bulk, lubricants, and mixture of
particles. As we understand, the added lubricant particles attach to
the surface of bulk particles due to cohesive interaction between the
two as given by Eq.~(\ref{cohesion}). The bulk particle interactions,
then, comprise contacts between the frictionless lubricant particles
coated on its surface in addition to direct contacts between
frictional (uncoated) surface of bulk particles. This reduces
effective friction between bulk particles causing them to flow past
one another with relative ease. The result is a faster flow with
enhanced inter-particle collisions and thereby an expanded flowing
layer [see Figs.~\ref{fig3}(a)-\ref{fig3}(d)].  The continual increase in lubricant
content progressively increases the coated surface area of the bulk
particles resulting in progressive reduction in friction and faster
flow. The flow rate achieves maximum value at $N_{r} = 300$ (see
Fig.~\ref{fig5}).  A further increase in $N_{r}$ is expected to
increase the coating area of the bulk particle. However, given only limited
bulk particle surface area for the coating, this also results in
several freely flowing lubricant particles not participating in
coating the bulk particles [see Figs.~\ref{fig3}(e)-\ref{fig3}(g)]. It is our
understanding that this excessive accumulation of lubricant particles
will increase damping of momentum in the system countering the
enhanced flow due to lubrication. The former effect seems to dominate
with continual increase in lubricant content resulting in progressive
reduction in the flow rate accompanied by increased volume fractions
(Fig.~\ref{fig4}) and compressed flowing layer [see
Figs.~\ref{fig3}(e)-\ref{fig3}(g)]. While this non-monotonic dependence of flow rate
on lubricant content looks to be in qualitative agreement with that
observed previously in
experiments,~\cite{ghodakeflow,RavindraparticleEffect} the results
exhibit certain nuanced differences as discussed next.

The flow rate of the particle mixture is higher than that of bulk
particles at the same value of $N_{r}$ even though its volume fraction is
higher. Second, maximum enhancement in the mixture flow rate in
simulations is observed to be around ten times that of the base
case. This is quite large compared to $1.5$ times enhancement observed
in experiments (not shown).  Furthermore, for $N_{r} > 300$, the decrease
in the flow rate of the mixture is observed to be gradual compared to the
steep decrease for bulk particles. The primary reason for these
observations is the spherical shape and larger size of lubricant
particles used in simulations. Apart from friction reduction at contact between bulk
particles, the larger-sized, spherical lubricant particle can also flow independently when unattached to bulk particles. The flow rate
of lubricant particles, thus, increases monotonically with increasing
$N_{r}$, with progressively increasing contribution from unattached
lubricant particles. Such independent flow rate of lubricant particles
contributes toward a higher flow rate of mixture. This behavior is
quite different from that observed in
experiments,~\cite{ghodakeflow,RavindraparticleEffect} wherein the
plate-like lubricant particles remain attached to the surface of bulk
particles at all times and cannot flow independently. The flow rate of
particle mixture in experiments is the same as the bulk particles, and it
gets enhanced only due to reduced friction between bulk particles at
contact.

\begin{figure}
  \includegraphics[scale=0.5]{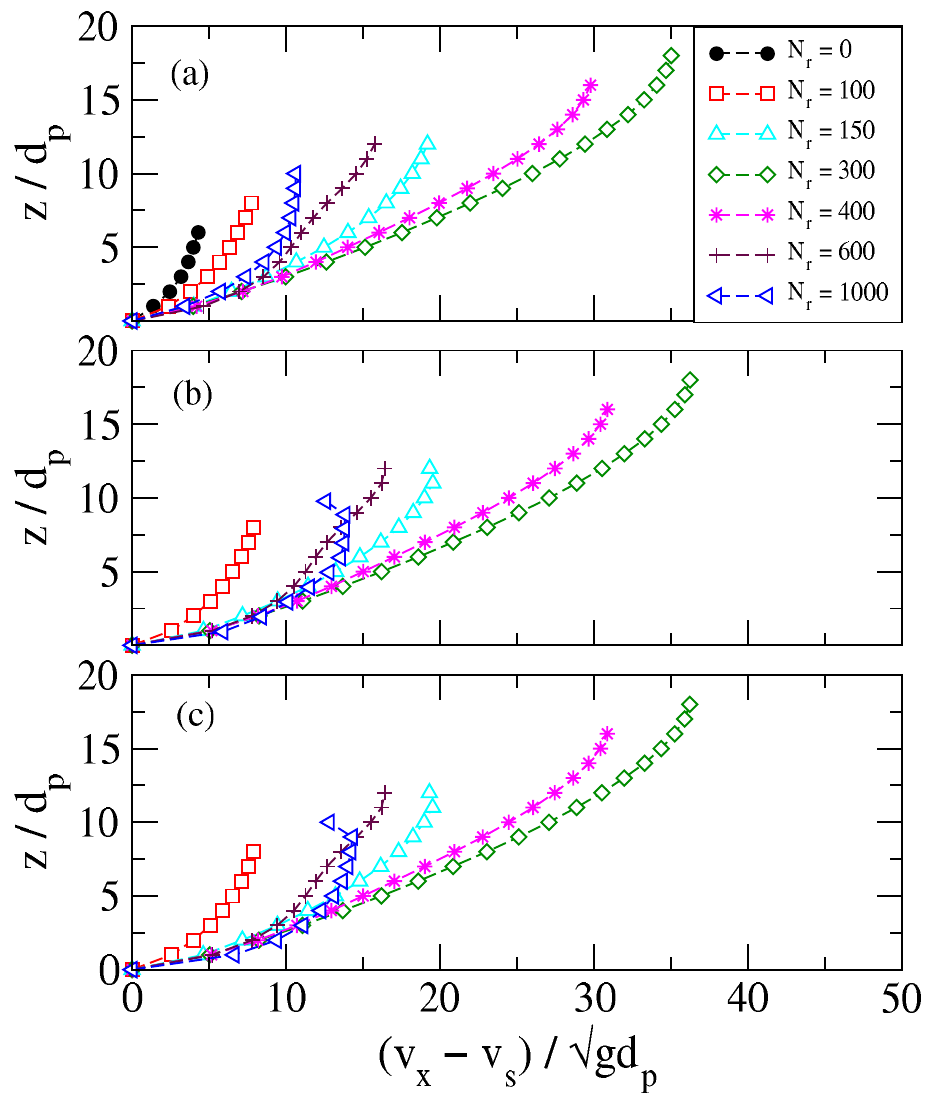}
  \caption{Variation of steady-state, normalized, flow direction average velocity $(v_{x} - v_{s}) / \sqrt{gd_{p}}$ profile with
    normalized distance from the chute bed ($z/d_{p}$) for
    different values of number ratio ($N_r$). Profiles are shown for (a) only bulk
    particles, (b) only lubricant particles, and (c) particle
    mixture. Error bars are smaller than the symbol size and hence are
    not shown to ensure clarity in the figure. See text for more details.}
  \label{fig6}
\end{figure}

The variation of steady-state, normalized, flow direction average velocity
[$(v_{x} - v_{s}) / \sqrt{gd_{p}}$] profile with normalized distance from the chute bed
($z / d_{p}$), for different values of $N_r$, is shown in Fig.~\ref{fig6} for bulk particles in (a),
for lubricant particles in (b) and for the mixture of bulk and
lubricant particles in (c).  All the profiles are plotted after
subtracting the slip velocity ($v_{s}$) from the actual velocity
values in the flowing layer. This ensures that the observed behavior
is only due to the interactions occurring between particles in the
flowing layer. The velocity profiles for all the cases, without
subtracting the bed slip velocity ($v_s$), are shown in supplementary material Figs. S2-S4.

\begin{figure}
  \includegraphics[scale=0.5]{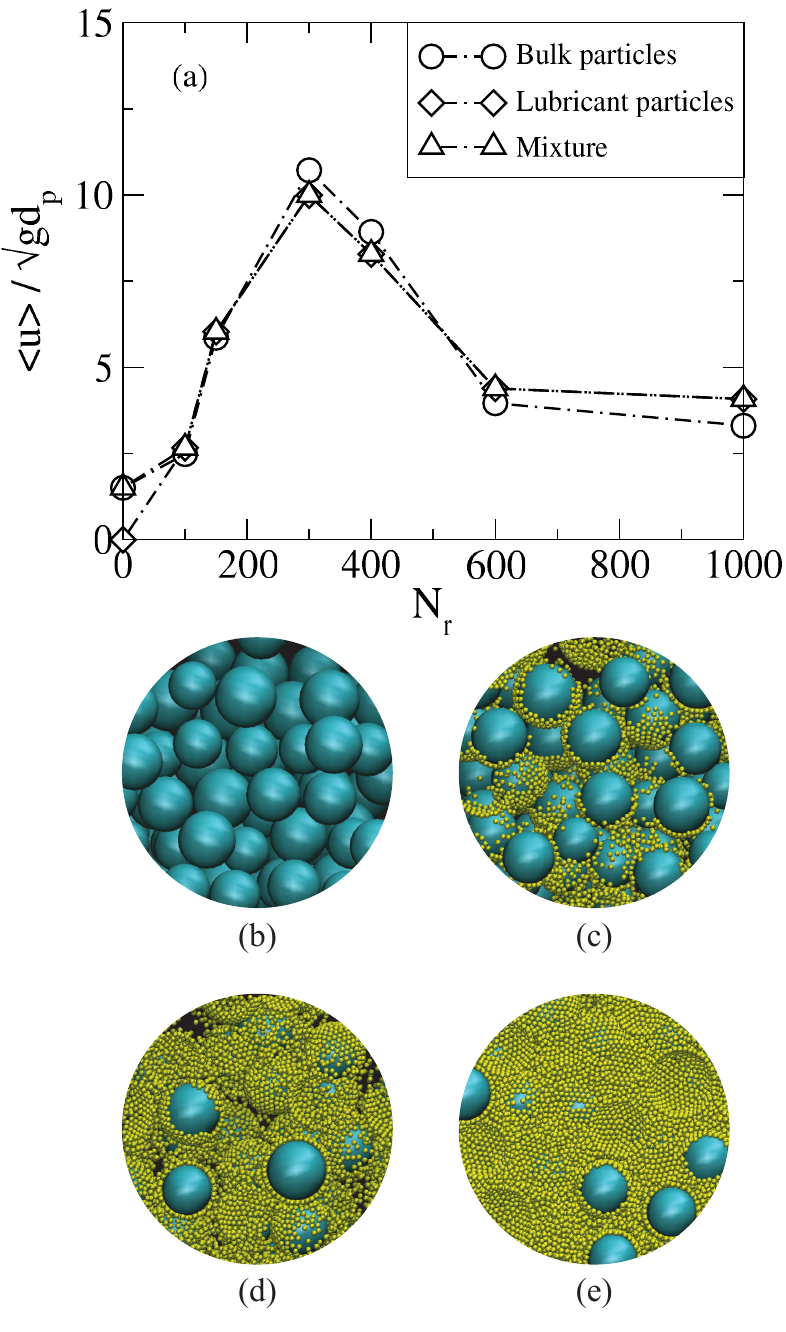}
  \caption{(a) Variation of steady-state, normalized, averaged rms velocity
    $\langle u \rangle / \sqrt{gd_{p}}$ of particles (bulk, lubricant, and mixture) in the flowing layer
  with the number ratio ($N_r$). Error bars are smaller than the symbol size and hence
    are not shown to ensure clarity in the figure. Magnified view of
    the instantaneous flow images for (b) $N_{r} = 0$, (c)
    $N_{r} = 100$, (d) $N_{r} = 300$ and (e) $N_{r} = 1000$ showing the
    particle network and overall densification. See text for more
    details.}
  \label{fig7}
\end{figure}

The profiles of velocity for all three cases, viz. only bulk
particles, only lubricant particles and particle mixture are nearly
the same qualitatively as well as quantitatively. This is not
surprising given that all of them are flowing together. However, the
flow rates ($\langle Q \rangle$) as shown in Fig.~\ref{fig5} are
different in each case given that the volume fractions
($\langle \phi \rangle$) are different. The velocity magnitude
progressively increases with increase upto $N_{r} = 300$, beyond which
it decreases continuously. The flowing layer thickness ($z / d_{p}$)
also exhibits non-monotonic dependence on $N_{r}$ in line with the
visual images shown in Fig.~\ref{fig3}. The increase in the velocity
magnitude is attributed to a reduction in the inter-particle friction
coefficient due to lubricant-coated bulk particles, while the
decrease is attributed to damping arising at increased volume
fractions, crowded particles in a compressed flowing layer.

The damping behavior at increased volume fractions can be ascertained by
calculating the root mean squared (rms) velocities ($u$) representing
the particle fluctuations or granular temperature. The crowded
environment at higher packing fractions is expected to damp out the
fluctuations due to repeated collisions through several particle
contacts, thereby leading to slower flow. The rms velocity ($u$) is
calculated as a single number for each value of $N_{r}$ number averaged over
all particles in the entire flowing layer, all three flow directions,
and over $1000$ time intervals. It is defined as
$u =\sqrt{[\langle c_{x}^{2} \rangle - \langle c_{x} \rangle
  ^{2}]+[\langle c_{y}^{2} \rangle - \langle c_{y} \rangle
  ^{2}]+[\langle c_{z}^{2} \rangle - \langle c_{z} \rangle
  ^{2}]}$. Here, $c_{x}$, $c_{y}$ and $c_{z}$ are the instantaneous
velocities of particles, respectively, in $x-$, $y-$ and $z-$
directions at every time interval. The variation of normalized rms
velocity ($u/\sqrt{g d_{p}}$) with $N_{r}$ for bulk, lubricant, and
  particle mixture is shown in Fig.~\ref{fig7}(a). The values are
  nearly the same for all three types across all values of
  $N_{r}$. The qualitative variation closely follows what is observed
  visually in Fig.~\ref{fig3}. The increased lubrication leads to a
  faster flow with an expanding layer resulting in increased fluctuations,
  which reach a maximum at $N_{r} = 300$,
  corresponding to a maximum in the flow rate and a minimum in the volume
  fraction. The volume fraction increases for a further increase in
  $N_{r}$ resulting in a higher volume fraction, increased crowding and
  damping leading to a reduction in fluctuations. Given that the flow of
  particles slows down at $N_{r} = 1000$, but does not stop completely, it leads to a slower reduction in the value of $u$. The magnified
  images shown in Figs.~\ref{fig7}(b)-\ref{fig7}(e) give a clear glimpse about the
  crowding and networking of particles for different values of $N_{r}$
  corroborating the explanation provided above.

\subsection{\label{parameter}Surface coverage and contact network analysis}

The images shown in Fig.~\ref{fig3} and all the subsequent discussion
presume increased coverage of the surface area of bulk particles by
lubricant particles for increasing $N_{r}$, which is quantified
next. The lubricant particle is considered to be adhered to a bulk
particle if the contact overlap between them is positive so that the
cohesive force acts, according to SJKR model.~\cite{CohesiveModels,
  coetzee2020sjkr} Considering that both lubricant and bulk
particles, are spherical in shape, a point contact is expected between
the two particles. The area covered (or coated) by the lubricant is,
then, defined as the area of its projection. For simplicity, this area
is considered to be that of a circle assuming contact without any
overlap. As the average overlap between particles is around $0.5$\% of
effective radius, the calculated surface coverage area may be
overestimated by a small amount.

The surface area coverage ($\sigma$) is defined as the ratio
of the total projected area of all the lubricant particles in contact with
the bulk particle to the total surface area of the bulk particle. Furthermore,
it is ensured that multiple contacts per lubricant particle are
considered, i.e., if a lubricant particle is in simultaneous contact
with two bulk particles, its coverage over both particles is
considered while calculating the average surface area. All contacts
between the bulk and lubricant particles were identified from the
particle positions at a particular time instant. A total of $1000$
such time instances, spread uniformly over the duration of steady-state regime, were considered, and the mean surface area coverage
$\langle \sigma \rangle$ was obtained as the average over all these time instances.

\begin{figure}
  \includegraphics[scale=0.45]{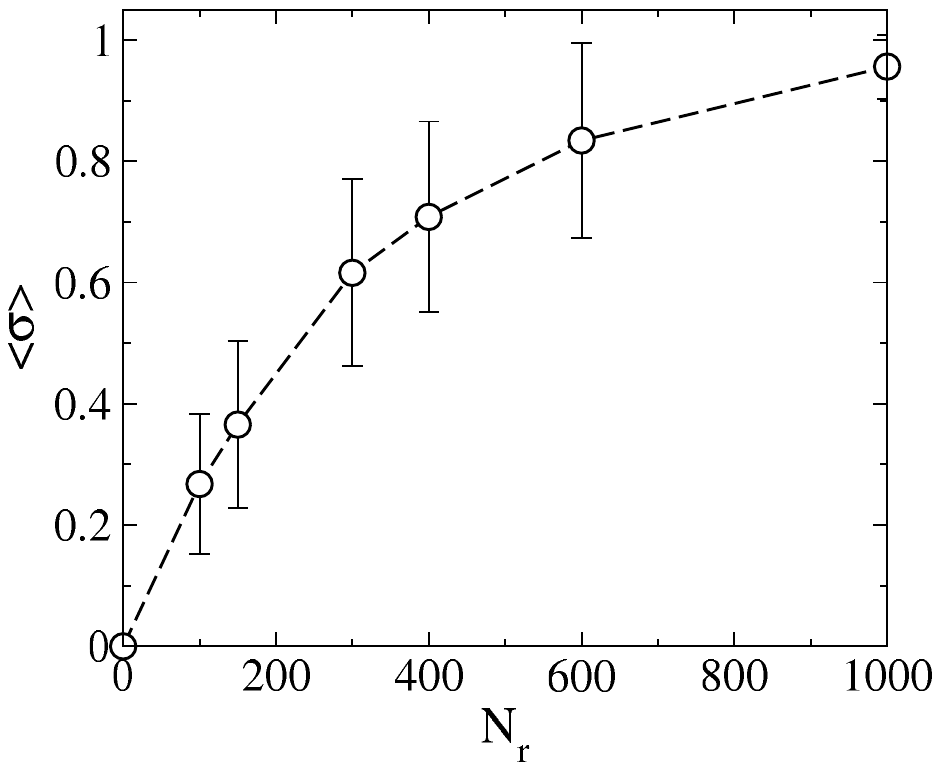}
  \caption{Variation of mean bulk particle surface area coverage
    ($\langle \sigma \rangle$) with the number ratio ($N_r$). The
    error bars represent spatial variations at any time instant and
    temporal variations over all time instants.}
  \label{fig8}
\end{figure}

The variation of $\langle \sigma \rangle$ with $N_{r}$ is shown in
Fig.~\ref{fig8}. The coverage increases rapidly in the initial phase,
as seen by the steepness of the curve. This is expected given a large
amount of surface area of bulk particles available to be coated. About
$60$\% bulk particle area is coated by $N_{r} = 300$, corresponding to
the flow expansion phase. Post $N_{r} = 300$, the rate of increase in surface coverage slows down. This slow-down is due to the reduction in
availability of bare area on the surface of bulk particle for higher
$N_{r}$. The near flattening of the curve by $N_{r} = 1000$ suggests
saturation to occur asymptotically at $0.95$. This may seem a bit
misleading, since a simple calculation using the total available surface
area of bulk particles and the total projected area of spherical
lubricant suggests complete coverage for $N_{r} = 400$ for a size ratio
of $10$. The discrepancy arises due to the fact that the surface of every
bulk particle is not completely coated by lubricant particles even
though the images in Fig.~\ref{fig3} seem to suggest so. We conjecture
the existence of volume exclusion effects that dissuade newer lubricant particles from connecting to bulk particles due to already coated lubricant
particles, though this could not be verified independently. This
effect is expected to accentuate at higher values of $N_{r}$. The
distributions of surface area coverage (not shown here) exhibit
distinct tails suggestive of variability of surface coverage across
all bulk particles, which reduces with increasing $N_{r}$ but does not
disappear even at $N_{r} = 1000$. The result is the asymptotic
approach of $\langle \sigma \rangle $ toward $1$ at higher value of
$N_{r}$.

\begin{figure}
  \includegraphics[scale=0.47]{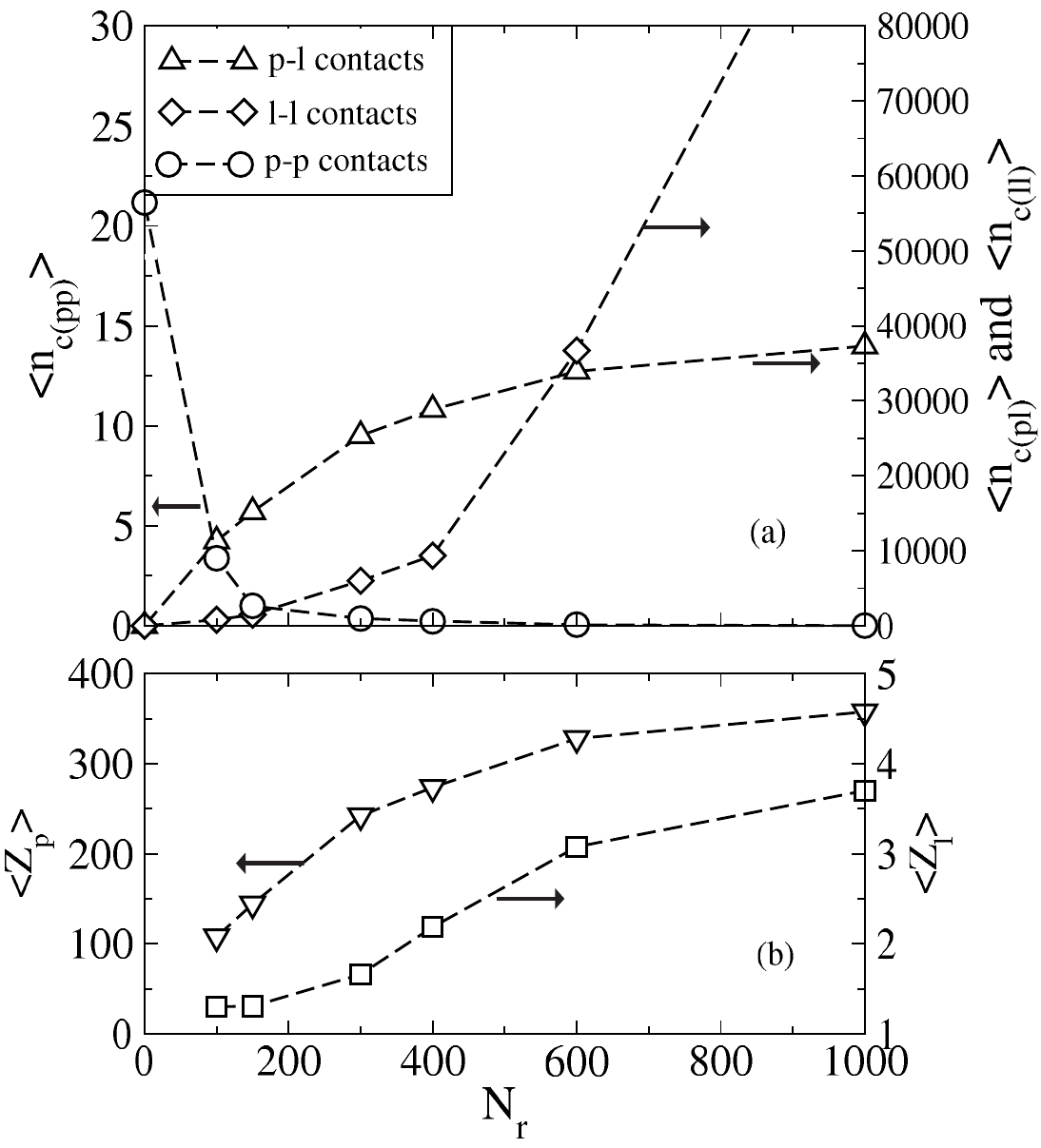}
  \caption{(a) Variation of average number of binary contacts with the
    number ratio ($N_{r}$). The y-axis on the left side denotes
    contacts between bulk particles ($\langle n_{c(pp)} \rangle$). The
    y-axis on the right side denotes contact between bulk and
    lubricant particles ($\langle n_{c(pl)} \rangle$) and between
    lubricant particles ($\langle n_{c(ll)} \rangle$). (b) Variation
    of average coordination number for bulk particles
    ($\langle Z_{p} \rangle$) and lubricant particles
    ($\langle Z_{l} \rangle$) with the number ratio ($N_{r}$). The arrows are provided on the curves
    for ease in understanding. Error bars are smaller than the symbol
    size and hence are not shown to ensure clarity in the figure.}
  \label{fig9}
\end{figure}

To further probe the structural information in the flowing layer, we
have examined inter-particle binary contacts and their variation with
an increase in the lubricant contents. Three types of contacts are
possible in the system, viz., between two bulk particles (p-p), bulk
and lubricant particles (p-l), and two lubricant particles (l-l). Any
two particles are considered to be in contact if the distance between
their centers is less than the sum of their radii, indicating
overlap. All such possible contacts are identified at a given time
instant within the steady-state flow regime. As for surface coverage
area measurements, a total of $1000$ independent time instants, spaced
uniformly within the steady-state zone, are considered, and the contact
data are number-averaged over all these time instants. The variation of
the mean binary contacts with lubricant content is shown in
Fig.~\ref{fig9}(a) for all three types of contacts mentioned above
($\langle n_{c(pp)} \rangle$, $\langle n_{c(pl)} \rangle$ and
$\langle n_{c(ll)} \rangle$).

\begin{figure*} 
\includegraphics[scale=0.55]{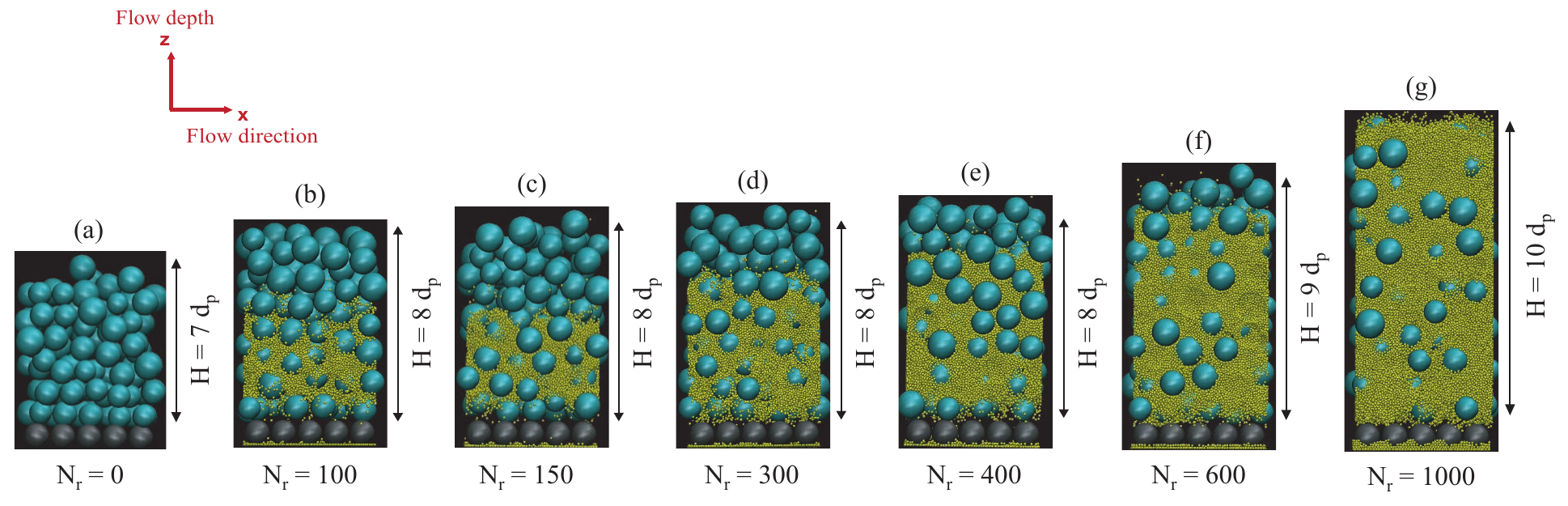}
  \caption{Instantaneous configurations
    (images) from simulations as viewed from the side at steady-state
    for different values of $N_{r}$. Bulk particles are represented as
    large blue spheres and lubricant particles as small yellow
    spheres. Cohesive interaction is absent between bulk and lubricant
    particles. Multimedia available online}
  \label{fig10}
\end{figure*}

\begin{figure}
  \includegraphics[scale=0.45]{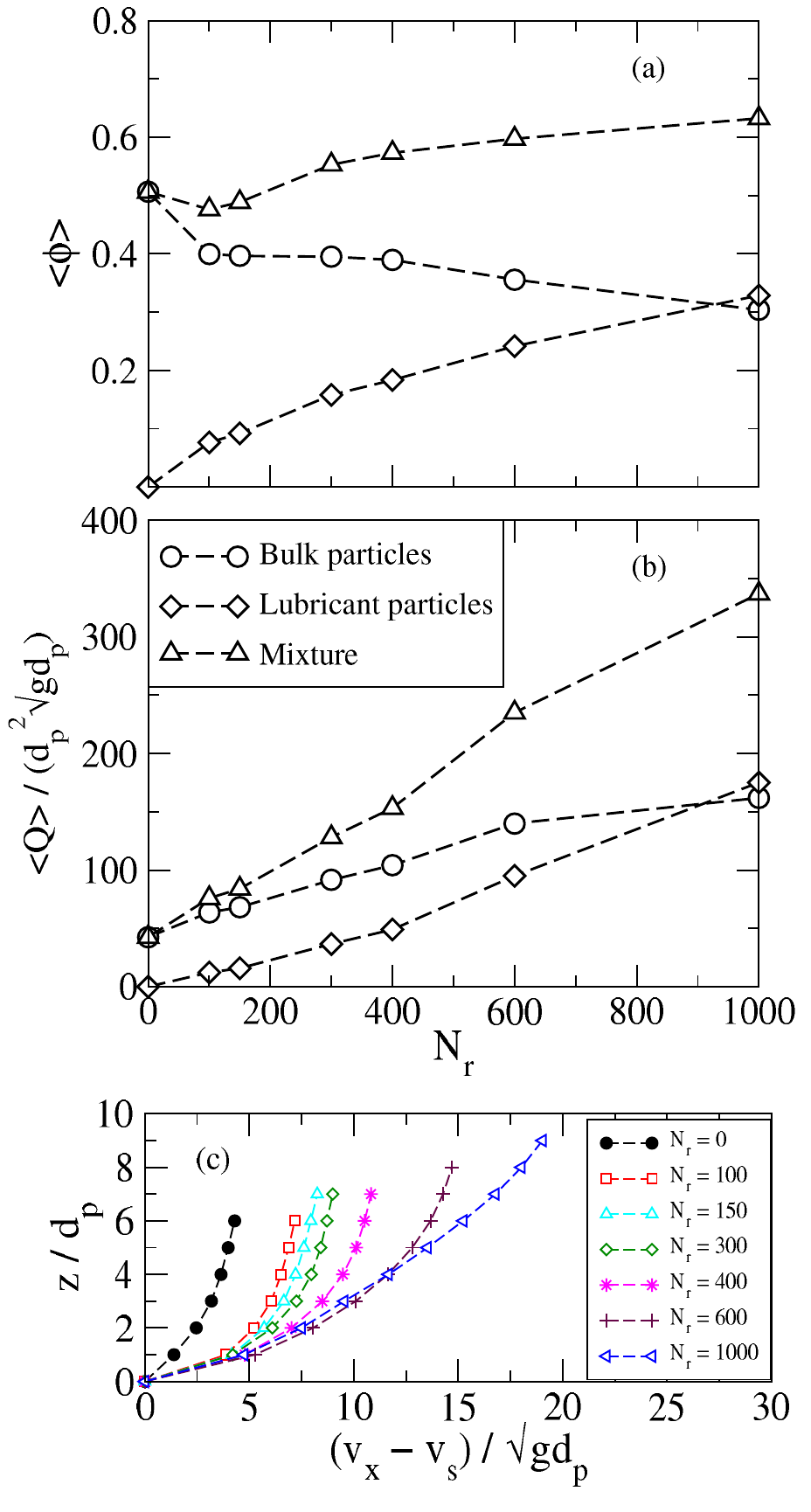}
  \caption{(a) Variation of steady-state mean volume fraction
    ($\langle \phi \rangle$) of particles (bulk, lubricant and
    mixture) in the flowing layer with the number ratio ($N_{r}$). (b) Variation of steady-state, normalized, mean flow rate
    ($\langle Q \rangle/d_{p}^{2}\sqrt{g d_{p}}$) of particles (bulk, lubricant, and
    mixture) in the flowing layer with the number ratio ($N_r$). (c) Variation of steady-state, normalized, flow direction average velocity $(v_{x} - v_{s}) / \sqrt{gd_{p}}$ profile for particle mixture with
    normalized distance from the chute bed ($z/d_{p}$) for different
    values of number ratio ($N_{r}$). Error bars are smaller than the symbol size and hence are
    not shown to ensure clarity.}
  \label{fig11}
\end{figure}

The mean values of all three types of contacts show monotonic
dependence on lubricant content.  The value of
$\langle n_{c(pp)} \rangle$ decreases rapidly by $N_{r} = 300$ with a simultaneous quick increase in the number of contacts between bulk and
lubricant particle ($\langle n_{c(pl)} \rangle$). Essentially, most of
the added lubricant particles coat the surface of bulk particles,
thereby reducing the number of contacts between two bulk particles
substantially.  The fall in $\langle n_{c(pp)} \rangle$ and rise in
$\langle n_{c(pl)} \rangle$ corresponds with flowing layer expansion
(see Fig.~\ref{fig3}) and faster flows (see Figs.~\ref{fig5} and
\ref{fig6}). Since most of the added lubricant is involved in
coating the surface of large particles, the number of contacts between
lubricant particles ($\langle n_{c(ll)} \rangle$) is lower and also
rises slowly. These can comprise particles floating around in the
system, unable to coat on the surface of bulk particles, and hence
remain in contact with other lubricant particles.

For a further increase in lubricant content, the number of contacts
between bulk particles approaches a near-zero value, while those between
bulk and lubricant particles reach saturation value by $N_{r} = 600$
[see the plateauing curve in Fig.~\ref{fig9}(a)].  This suggests that
the entire surface of all the bulk particles in the system is nearly
coated by lubricant particles. Thus, no further contact is possible
between lubricant and bulk particles or between bare surfaces of two
bulk particles. The continual addition of lubricant particles
continues to fill the void space available in the system. This leads
to an increase in the number of contacts between lubricant particles
($\langle n_{c(ll)} \rangle$), which exhibits a rapid rise for
$N_{r} > 400$. This corresponds to increased volume fraction (see
Fig.~\ref{fig4}), compressed flowing layer [see Fig.~\ref{fig3}(g)],
reduced flow rate (see Fig.~\ref{fig5}), and reduced velocity (see
Fig.~\ref{fig6}, blue symbols). In principle, the number of contacts
between lubricant ($\langle n_{c(ll)} \rangle$) particles can continue
to increase indefinitely with increasing lubricant content (no
saturation expected as in the other two cases). However, at a certain very
high lubricant concentration, the total volume fraction may increase
to such an extent that would lead to complete stoppage of flow (not
calculated here) due to severe damping of the flow momentum.

Figure~\ref{fig9}(b) shows the variation of coordination number for bulk
($Z_{p}$) and lubricant ($Z_{l}$) particles with $N_{r}$. The
coordination number is defined as the number of particles (bulk as
well as lubricant) in contact with either bulk ($Z_{p}$) or lubricant
($Z_{l}$) particle. The contact definition is the same as used for
surface area coverage calculations. For both, lubricant and bulk
particles, the coordination number increases quickly upto
$N_{r} = 400$ before slowing down at higher values of $N_{r}$. The
initial quick increase is attributed to more available surface area
for contact, while the near plateauing at higher values of $N_{r}$
indicates progressive decrease in available surface area. The value of
$Z_{p}$ asymptotically approaches $400$ corresponding to
complete coverage by lubricant particles with no contact with
another bulk particle. This is synchronous with the asymptotic
approach of $\langle \sigma \rangle$ toward $1$ shown in
Fig.~\ref{fig8}. The value of $Z_{l}$ also plateau at higher values of
$N_{r}$ though the asymptotic value is not clear given that the
lubricant particle may be in contact with both bulk and other lubricant particles.

\subsection{\label{parameter}Role of cohesion}

The observations shown above and the subsequent discussions
correspond to the lubricant particles coating on the surface of bulk
particles due to cohesive bonding.  The obvious question is whether
cohesion between bulk and lubricant particles is necessary or if merely
the particle mixture without any cohesive bonding will produce a similar
effect. To ascertain this issue, simulations were repeated for the
same conditions as for the cases shown in Fig.~\ref{fig3}, but without
incorporating any cohesion between lubricant and bulk particles. The
resulting instantaneous particle configurations, during the steady-state flow, are shown in Fig.~\ref{fig10} (Multimedia view) for various values of $N_{r}$.  Certain qualitative
differences are evident with respect to the images shown in
Fig.~\ref{fig3} which we discuss next.

Unlike the non-monotony observed in Fig.~\ref{fig5}, over here the
flow expands monotonically with increasing lubricant
content. Moreover, the expansion commensurates approximately with the
volume added due to lubricant particles. For instance, the total
volume of particles doubles at $N_{r} = 1000$ and the flowing layer
expands about $1.6$ times, which is close to $2$ but exactly not the
same. This is expected given that small particles will try and
maximize the usage of voids between bulk particles by inserting
themselves in those spaces, thereby increasing the packing fraction
and in the process expanding the flowing layer. Certainly, the flow
layer expansion is not as exaggerated as observed for cohesive
systems. Second, the mixture of particles separate in two layers
similar to that observed in sieving/percolation segregation. This
tendency is high for low values of $N_{r}$ and diminishes for higher
values of $N_{r}$. The reason for the latter is the sufficiently
densified and less fluidized flow, which may prevent the separation of
particles by the sieving/percolation process. This difficulty in
separation seems to increase progressively with increase in
$N_{r}$. Also, it is to be noted that the separation of particles in
layers, if it happens at all, occurs within a short time of flow and is
retained till steady state is achieved. The particle configuration
shown in Fig.~\ref{fig10} will, thereby, remain the same even if the
flow is continued for a longer duration. 

The observations noted above are quantified to obtain the variation of
mean volume fraction ($\langle \phi \rangle$), flow rate
($\langle Q \rangle$) and flow direction average velocity
($v_{x} - v_{s}$) with $N_{r}$. The data are shown in
Fig.~\ref{fig11} for all three quantities. First and foremost, all the
quantities exhibit monotonic variation with $N_{r}$ unlike the
non-monotonous behavior observed with coated lubricant particles. The
volume fraction of bulk particles decreases continuously while that of
lubricant particle increases continuously with $N_{r}$ [see
Fig.~\ref{fig11}(a)]. The mixture volume fraction remains nearly flat
for all values of $N_{r}$. The flow rate for bulk, lubricant and
particle mixture increases continuously with $N_{r}$, with the flow
rate of mixture approximately the sum of the other two [see
Fig.~\ref{fig11}(b)]. This is simply the consequence of segregated
layers flowing together, as observed in Fig.~\ref{fig10}. Quite
obviously, the velocity magnitude and flowing layer thickness increase
monotonically with increasing $N_{r}$ [see
Fig.~\ref{fig11}(c)]. Overall,the observations suggest that the flow, in 
the absence of cohesion, seems to be fundamentally different from that
taking place in the presence of cohesion, the latter being a necessity
to enable a lubrication effect. At the same time, we expect (though not
investigated) that the cohesion intensity should be just enough to
allow lubrication and cannot be too high, which otherwise could lead
to clustering of the flow and change the fundamentals of the flow
altogether.

\section{\label{conclusion}CONCLUSION}

The gravity-driven flow of large-sized granular particles, mixed with a prescribed amount of small-sized lubricant particles, on an inclined
plane is investigated using discrete element method (DEM)
simulations. The simulations are carried out using periodic boundaries
in the flow and transverse directions so that they resemble flow behavior away from side walls as well as away from the entry and exit regions of the chute. A cohesive force is enabled between large and small particles
whereby the latter coat the surface of the former. Given the near
frictionless surface of the small particles, their coating on the
surface of bulk particles tends to reduce the contact between
frictional larger particles while increasing contact between smaller
particles and consequently reducing the effective friction between two
bulk particles. This behavior resembles lubrication phenomena as they have been studied previously through
experiments.~\cite{ghodakeflow,morineffect}

The simulations show that the mean velocity, flow rate and layer
thickness increase to a maximum with increasing lubricant
content. This occurs due to reduced effective friction and presumably
enhanced collision from increased agitation. Post the maximum
achievable flow rate improvement, a further increase in lubricant
content reduces the mean velocity, flow rate, and layer thickness. The
overall behavior is found to be qualitatively similar to that observed
in experiments and is characterized in terms of packing fraction, flow
rates, and inter-particle contacts. However, a few differences from
experiments persists, particularly the mechanism for decreased flow
rate at higher lubricant content. In simulations, the decrease is
attributed to a substantial increase in the volume of lubricant
particles thereby causing densification of the system. In experiments,
the decrease is attributed to increased friction or damping with the
lubricant particles remaining attached to bulk particles throughout
with negligible volume. These differences are observed due to
approximating the actual plate-like lubricant particle as a sphere.

Using spherical particles as a lubricant simplifies interaction between particles at a single contact point, instead of planar and several
oblique contacts of a plate-like lubricant with a bulk particle
surface. Effectively, this reduces the number of computations and
simulation duration significantly while allowing for the use of a
simple and well-established interaction model. The simplification,
however, cannot replicate the true lubrication mechanism by spreading
and stretching of plate-like lubricant in the form of a thin film on
the surface of bulk particles. While this can be obviated by using
tiny spherical particles, a constraint is posed due to the increased number
of particles, leading to a higher computational cost. Notwithstanding
these limitations, the simulations do provide a reasonably good
approximation of the overall experimental behavior while providing an
adequate explanation for it.

At the same time, we believe that the simulation results provide a
newer, interesting particle assembly, bi-disperse particles with
induced cohesion. The presence of solid lubricant, in a way, acts as
an external agent to alter effective friction in the particle assembly
and resulting shear rate. This behavior resonates quite well with the
well known $\mu (I)$ rheology~\cite{gdrmididense,jop2005} which
relates effective friction in a particle assembly with the imposed
shear rate and has been shown to predict granular flows in various
systems quite well. It would be interesting to investigate whether
this rheology can be seamlessly implemented for the study over here or
the presence of lubricant and its cohesiveness needs to be explicitly
accounted for in the constitutive equation.

The results from this work also offer valuable implications for
optimizing lubricant concentration in pharmaceutical and other
industries (e.g., mining) where powder processing is critical. The
critical factor in these processes is the aspherical bulk particle,
the size of which is typically an order of magnitude lower than used
in experiments. This induces additional cohesion between bulk
particles which is not considered over here and can lead to a more
complicated flow. Second, the applicability of the results over here
also need to consider the equipment sizes in industry, which can pose
its own scalability issues. Thus, while the results obtained here are
interesting and provide an initial attempt toward understanding this
complex process, much further work is needed for ascertaining the
carryover of this phenomenon to more complicated flows.

\section*{ACKNOWLEDGEMENTS}

The authors acknowledge several fruitful and insightful discussions
with Dr. Pankaj Doshi, which were of significant help in this work.
A.V.O. gratefully acknowledges the financial support from Science \&
Engineering Research Board, India (Grant No. CRG/2019/000423). S.V.C.
acknowledges the Council of Scientific and Industrial Research (CSIR),
India for the GATE fellowship. The ``CSIR centralized HPC, AI \& ML
Platform (CHAMP) facility'' provided by CSIR-4PI, Bengaluru; the ``PARAM Brahma Facility'' at the IISER, Pune; and the ``Einstein cluster
facility'' at CSIR-NCL, Pune are gratefully acknowledged for providing
necessary computational resources.

\bibliography{citations}

\end{document}